\documentclass[a4paper,12pt]{article} % only 10 (default), 11 and 12 pt are available

\usepackage[utf8]{inputenc} % to be able to use non-English characters
\usepackage{amsmath}  % improve math presentation
\usepackage{amssymb}
\usepackage{xcolor}
\usepackage{float}
\usepackage{graphicx}
\usepackage{chemformula}
\usepackage{fancyhdr}

\usepackage{caption} % captioning package of wonders
\usepackage{url} % to incorporate clickable links
\usepackage{cite} % takes care of citations
\usepackage[final]{hyperref} % adds hyper links inside the generated PDF file
\usepackage{multicol}

\usepackage{newfloat} % for defining new float in environments (e.g: figure and tables are floats)
\usepackage{graphicx} % takes care of graphic including machinery
\usepackage{tabularx} % extra features for tabular environment
\usepackage{array} % provides 'programmable' tables (cells can be tweaked much more)
\usepackage[export]{adjustbox} % to include boxed content other than figures (e.g: boxed equations)
\usepackage{wrapfig} % to make figures wrap around text
\usepackage{subcaption} % to make sub-figures inside of figures

\usepackage{parskip}

\usepackage[margin=2cm,a4paper]{geometry} % tweaks margins
\numberwithin{equation}{section} % equations are named according to their section (e.g: 2.1)
\numberwithin{figure}{section} % figures are named according to their section (e.g: 3.4)
\hypersetup{
    colorlinks=true,      % false: boxed links; true: colored links
    linkcolor=black,      % color of internal links
    citecolor=blue,       % color of links to bibliography
    filecolor=magenta,    % color of file links
    urlcolor=red          % color of website links
}

\usepackage{lipsum} % DELETE THIS AND BELOW \lipsum[]'s AND YOU'RE GOOD TO GO

\newcommand{\hwcourse}{\text{Magnetoplasmadynamic Thrusters}} % Title of your document

\newcommand{\hwname}{\text{Magnetohydrodynamics }} % Name of your study name
\newcommand{\hwdetails}{ \text{A Brief Overlook on} \\ 
                            \text{Magnetoplasmadynamic Thrusters} \\ % Details about your study
                            \text{and Literature Review} }

\newcommand{\hwauthor}{ Egemen Gover} % Your name or your group's names
\newcommand{\HRule}{\rule{\linewidth}{0.5mm}} % line widths in the cover page

\begin{document}

% COVER PAGE IS COMPILED HERE
\begin{titlepage}

\begin{center} % Center remainder of the page
% LOGO SECTION
\includegraphics[width = 8cm]{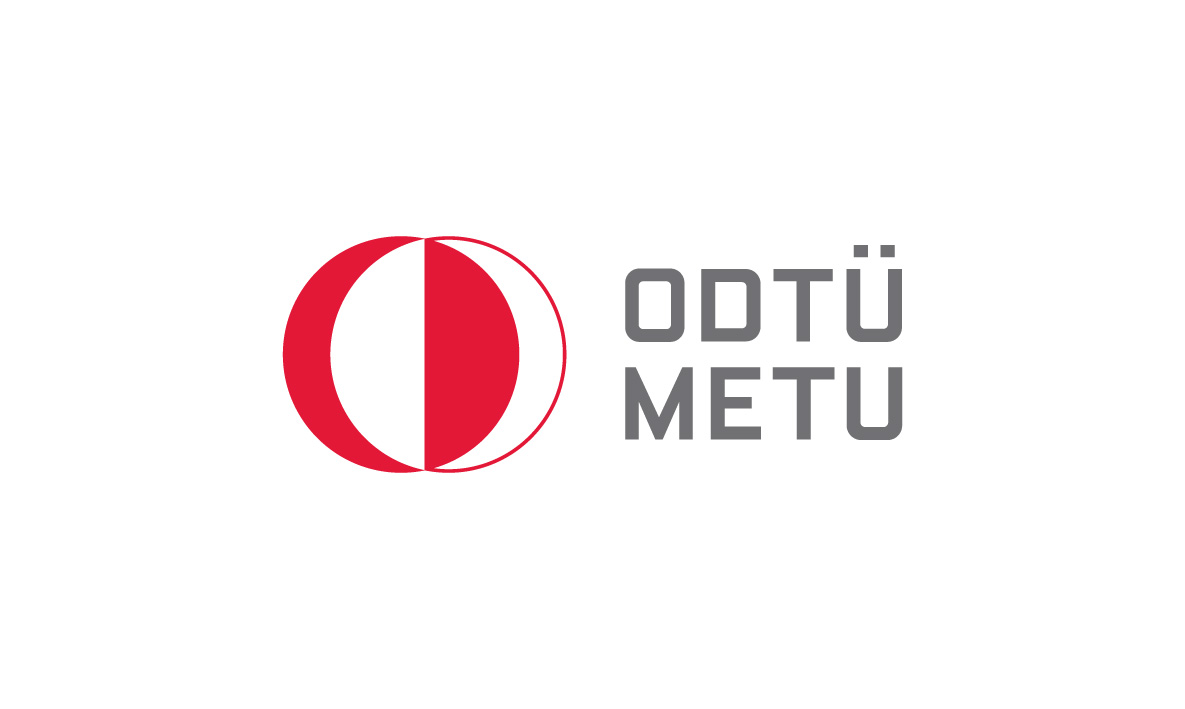}

% HEADING SECTIONS
\textsc{\LARGE Middle East Technical University}\\[.5cm]
\textsc{\Large Physics Department}\\[1.5cm]

% TITLE SECTION
\HRule \\[0.4cm]
{ \huge \bfseries \hwcourse}\\ \vspace{.5cm} 
{ \large \bfseries \hwname}\\ \vspace{.5cm} 
{ \hwdetails}\\ \vspace{.5cm}
\HRule \\[1.5cm]
\end{center} 

% AUTHOR SECTION
\begin{center} % left oriented author section
    \large 
    \textit{Written By:}\\
    \hwauthor% Your name
\end{center}
\vspace{5cm}
\makeatletter
Date: \@date 
\vfill % Fill the rest of the page with white space
\makeatother
\end{titlepage}

%++++++++++++++++++++++++++++++++++++++++++++++++++++++++++++++++++++++++++++++++

\newpage

\tableofcontents

\newpage

\section{Introduction}

As advancements in deep space exploration are accelerating, it is of utmost importance that spacecraft have reliable, efficient thruster technology. While we currently have high-thrust combustible engines, for missions concerning high specific impulse, chemical engines are practically useless, as their burn duration is considerably less then their electric thruster counterparts. In addition to the absence of high specific impulse, another drawback is that the use of combustion engines on spacecraft requires large volumes of combustible material embedded within the craft itself. This in turn increases the overall mass, dramatically decreasing cargo space, and hence increasing the thrust needed to accelerate even more mass .Space technologies need to be affordable, hence the concept of decreasing the overall mass further decreases the cost, which is ideal. To overcome this problem, a new propulsion technique called \textit{electric propulsion} was first demonstrated in the 1960s. Although this can be regarded as a breakthrough in propulsion technology, we still cannot put the chemical engine on the shelf, as we are still reliant on chemical engines for take-offs from Earth due to inadequacy of EPs in providing the thrust needed to overcome gravity.

\subsection{Specific Impulse}

What really puts forward the electric propulsion is specific impulse. Specific impulse is the direct measurement of efficiency of thruster. As such, it defines how efficiently the propellant's mass is used to produce thrust. Higher specific impulse implies that the change of velocity  \(\Delta v\) provided by the engine requires less fuel compared to a lower specific impulse thruster. Usually, thrust and specific impulse are inversely proportional, meaning higher SI compromises thrust.
\\~\\
Specific impulse ($I_{sp}$) is defined as:
\\
\begin{equation} \label{eq1}
    I_{sp} = \frac{Thrust}{\textit{Mass flow rate of the propellant}} = \frac{T}{\hphantom{^2}\dot{\omega_{f}}}
\end{equation}

Thrust has units of force. It is the main pushing force of engine, generating the necessary push to take-off, relocate and navigate. General idea of a propulsion system is that using Newton's third law, if a gas is accelerated towards a direction, the craft will react by moving in opposite.
As though when we think of an electric propulsion system, one would be tempted to think there will be no fuel in the system, which is entirely wrong. It is important to note that we will still have a propellant, mostly inert gas in EP case, to push the rocket.

Rearranging \ref{eq1}:

\begin{equation}
     \dot{m}v_{e} = \dot{\omega_{f}}I_{sp}
\end{equation}
\begin{equation}
     I_{sp} = \frac{v_e}{g}
\end{equation}
where $v_{e}$ is the exhaust velocity.

The beauty of electric propulsion is that in some various engine designs, the exhaust velocity can be altered just by changing the electromagnetic field, therefore resulting in a variable specific impulse (i.e VASIMR). This implies that adjustable power input is independent of mass flow rate. Another important parameter for thrusters is the thrust efficiency $\eta$ which is given by:

\begin{equation}
\eta = \frac{F^2}{2\dot{m}JV_d}
\end{equation}

In general the thrust efficiency has a linearly proportional relation with specific impulse. This implies that lighter elements provide greater specific impulse with greater thrust efficiency.

\begin{table}[h]
\centering
\begin{tabular}{|c|c|c|c|c|c|}
\hline
\textbf{Propellant} & $\mathbf{\dot{m} \ [kg/s}]$ & $\mathbf{J \ [Ampere]}$ & $\mathbf{B \ [Tesla]}$ & $\mathbf{V_d \ [Volt]}$ & $\mathbf{F \ [Newton]}$ \\
\hline
H$_2$ & $9.0 \times 10^{-7}$ & 200 & 0.05 & 49 & 0.028 \\
      & $9.0 \times 10^{-7}$ & 200 & 0.075 & 39 & 0.042 \\
      & $9.0 \times 10^{-7}$ & 200 & 0.10 & 36 & 0.050 \\
      & $1.2 \times 10^{-6}$ & 200 & 0.075 & 39 & 0.033 \\
\hline
He    & $1.8 \times 10^{-6}$ & 200 & 0.025 & 32 & 0.014 \\
      & $1.8 \times 10^{-6}$ & 200 & 0.05 & 31 & 0.029 \\
      & $1.8 \times 10^{-6}$ & 200 & 0.075 & 35 & 0.074 \\
      & $2.4 \times 10^{-6}$ & 200 & 0.075 & 26 & 0.048 \\
\hline
N$_2$ & $8.4 \times 10^{-6}$ & 200 & 0.25 & 32 & 0.071 \\
\hline
Ar    & $9.0 \times 10^{-6}$ & 200 & 0.10 & 18 & 0.063 \\
      & $9.0 \times 10^{-6}$ & 85 & 0.25 & 36 & 0.112 \\
      & $9.0 \times 10^{-6}$ & 110 & 0.25 & 38 & 0.123 \\
      & $9.0 \times 10^{-6}$ & 200 & 0.25 & 20 & 0.097 \\
\hline
\end{tabular}
\caption{Experimental operating conditions and their resulting thrust for typical propellants used in electric propulsion systems\cite{sasoh1992electromagnetic}.}
\end{table}

\subsection{Types of Ion Electric Propulsion}

In contrast to traditional chemical rocket engines, which fire hot gas to make thrust,  electric propulsion uses electromagnetic forces to push gas out of nozzle. Type of EPs are as follows:
\\~\\
\large{\textbf{Ion and Plasma Drives}}
\paragraph{Electrostatic:}
\normalsize{Thrust is obtained by Coulomb forces. Static electric field is the main driving force.} Example thrusters are as follows:
\begin{itemize}
    \item Hall-Effect Thruster
    \item Colloid Ion Thruster
    \item Nano-particle Field Extraction Thruster
\end{itemize}
So far, this type of engine has been tested in space environment (TRL 9). Unlike electrothermal thrusters, it uses heavier gases, such as \ch{Xe}. The thurst levels are relatively low. Exhaust velocities are of 70-90 km/s with thrust efficiency of 60-80\%

\paragraph{Electrothermal:}

\normalsize{Such engines use electromagnetic fields to create plasma. As the word  thermal implies,  these fields further heat up the plasma, where in turn the thermal energy converted into kinetic energy with the help of nozzle or additional fields.}
\begin{itemize}
    \item Variable specific impulse magnetoplasma rocket (VASIMR)
   
    \item Microwave Thruster
    \item Resistojet
\end{itemize}
Such thrusters have:
\begin{itemize}
    \item High trust levels which are up to 20 km/s    
    \item Light gases as propellant(H, He, \ch{NH3})
    \item Efficiency in the order of 20\% to 50\% 

\end{itemize}
\paragraph{Electromagnetic:}
\normalsize{Similar to electrothermal engines, but this time ions are accelerated with the help of Lorentz force }
\begin{itemize}
    \item Magnetoplasmadynamic Thruster
    \item Pulsed plasma thruster
    \item Pulsed inductive thruster
    \item Magnetic field oscillating amplified thruster
    \item Helicon Double Layer Thruster
    \item Electrodeless Plasma Thruster

\end{itemize}
Such thrusters have:
\begin{itemize}
    \item Medium thrust levels, ranging from 10-70 km/s
    \item Various propellants (either inert or reactive) \ch{H2}, He, Ar, Kr, Xe, Li
    \item Adjustable power range, due to the nature of applied magnetic and electric field.
\end{itemize}

\large{\textbf{Non-Ion Drives}}

\normalsize{\textbf{Photonic}: Thrust is based on the momentum carried by photons, transferred onto the craft.}

\begin{table}[h!]
    \centering
    $\begin{array}{lccccc}
\hline \text { Technology } & T[\mathrm{mN}] & P_{r e q}[\mathrm{~kW}] & \eta_t[\%] & I_{s p}[\mathrm{~s}] & \text { Propellants } \\
\hline \text { Resistojet } & 250-300 & 0.5-1.5 & 65-90 & 200-350 & \mathrm{H}_2, \mathrm{NH}_3, \mathrm{~N}_2 \mathrm{H}_4 \\
\text { Arcjets } & 200-1000 & 0.3-30 & 30-50 & 400-1000 & \mathrm{H}_2, \mathrm{NH}_3, \mathrm{~N}_2 \mathrm{H}_4, \mathrm{~N}_2 \\
\text { Ion Thruster } & 0.01-500 & 1-7 & 60-80 & 1500-10000 & \mathrm{Ar}, \mathrm{Kr}, \mathrm{Xe}, \mathrm{Bi} \\
\text { Hall Thruster } & 0.01-2000 & 1-10 & 50-55 & 1500-2000 & \mathrm{Ar}, \mathrm{Xe} \\
\text { Helicon Thruster } & 1-1000 & 0.05-50 & 10-40 & 500-2000 & \mathrm{Ar}, \mathrm{Xe}, \mathrm{Kr}, \mathrm{N}_2, \mathrm{H}_2 \\
\text { SF-MPD Thruster } & 0.001-50000 & 100-4000 & \leq 40 & 2000-5000 & \mathrm{Ar}, \mathrm{Xe}, \mathrm{Li}, \mathrm{H}_2 \\
\hline
\end{array}$
    \caption{Range of parameters of some typical electric propulsion systems where T is thrust, P is power required, $\eta_{t}$ is thruster efficiency, $I_{sp}$ is specific impulse. \cite{alvaro_thesis}}
    \label{tab:my_label}
\end{table}

\paragraph{In the following sections, we will work on magnetoplasmadynamic thrusters (also known as  Lorentz Force Accelerator (LFA), which is a specific type of steady state electromagnetic thrusters.}

\begin{figure}
    \centering
    \includegraphics[scale=0.6]{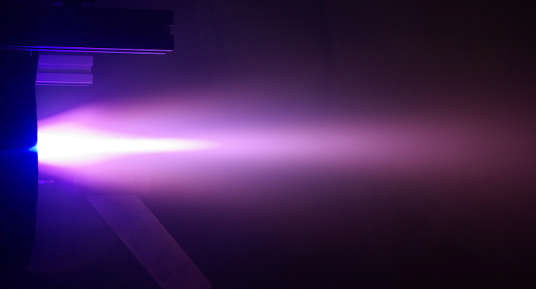}
    \caption{IRS Applied-Field MPD Thruster in testing} \cite{IRS_MPD}
    \label{IRS_MPDl}
\end{figure}

\section{Working Principles}
\subsection{The Working Principle Behind MPD - Lorentz Force}
Before we get into the nature of MPD thrusters, we shall start by analyzing the physics in first principles, meaning microscopic analysis shall be our first topic.  To do so, some simplifications, or rather assumptions need to be made.
\begin{itemize}
    \item The medium shall be assumed as plasma. It will consist of only electrons and single-charged ions, where each of the number densities are approximately the same ($n_{e}\approx n_{i}=n$) (quasi-neutrality)
    \item Momentum storing capability of electron fluid will be neglected, as the presence of heavy ions will more dominantly get into play. \cite{magnetoplasmadynamics}
\end{itemize}
For an MPD thruster to work, plasma is required to apply the following necessary formulas.  The creation of plasma inside the chamber is provided by the traversing electrons. As the cathode ejects prompt electrons, the gas tank of the engine activates and ejects gas (typically inert) into the chamber. Electrons colliding with this gas is responsible for ionization, and hence creation of ions takes place.

Now we can write the momentum conservation equations as\cite{magnetoplasmadynamics}:
\begin{equation}
m_{\mathrm{i}} n \frac{\mathrm{d} \boldsymbol{u}_i}{\mathrm{~d} t}=n e\left(\boldsymbol{E}+\boldsymbol{u}_{\mathrm{i}} \times \boldsymbol{B}\right)-\nabla p_{\mathrm{i}}+P_{\mathrm{ie}}
\end{equation}
\begin{equation} \label{lorentz-eq}
0=-n e\left(\boldsymbol{E}+\boldsymbol{u}_{\mathrm{e}} \times \boldsymbol{B}\right)-\nabla p_{\mathrm{e}}+P_{\mathrm{ie}}
\end{equation}

where $m_{\mathrm{i}}$ is the mass of ion,  ${u}_{\mathrm{e}}$ and ${u}_{\mathrm{i}}$ are electron and ion velocities, \textbf{\textit{E}} and \textbf{\textit{B}} are electric and magnetic induction, $p_{\mathrm{e}}$ and $p_{\mathrm{i}}$ are electron and ion pressures, $P_{\mathrm{ie}}$ is the momentum gain of ions due to collisions with electrons. Equation \ref{lorentz-eq} is crucial to understanding how MPD thrusters use \textit{Lorentz force} as their basis working principle. The collision terms we have included so far only accounts for the intrinsic collisions happening within the plasma itself (more generally, electrons and ions). Collisions between the charged particles fall in a category known as Coulomb (or glancing) collisions, which will be studied in more detail in the following sections.

Continuing with our idealization, we can assume that for intrinsic collisions not involving the nozzle or the chamber itself (that is, the inelastic collisions happening between the walls of thruster and the plasma) we can assume the momentum gain of electrons due to collision with ions is just the negative of momentum gain of ions due to collisions with electrons, and vice versa.
\begin{equation}
    \textbf{P}_{ie} = -\textbf{P}_{ei}
\end{equation}

Substituting this into equation \ref{lorentz-eq} we get

\begin{equation}
m_{\mathrm{i}} n \frac{\mathrm{d} \boldsymbol{u}_i}{\mathrm{~d} t}=n e\left(\boldsymbol{E}+\boldsymbol{u}_{\mathrm{i}} \times \boldsymbol{B}\right)-\nabla p_{\mathrm{i}}+n e\left(\boldsymbol{E}+\boldsymbol{u}_{\mathrm{e}} \times \boldsymbol{B}\right)-\nabla p_{\mathrm{e}}
\end{equation}

where we can further define the electron and ion pressure as:

\begin{equation}
    \nabla{p} = \nabla{p_{e}} + \nabla{p_{i}}  
    \qquad \rho = m_{n}n 
\end{equation}

Overall we have:
\begin{equation}
    \rho \frac{\mathrm{d} \boldsymbol{u}_{\mathrm{i}}}{\mathrm{d} t}=n e\left(\boldsymbol{u}_{\mathrm{i}}-\boldsymbol{u}_{\mathrm{e}}\right) \times \boldsymbol{B}-\nabla p=\boldsymbol{j} \times \boldsymbol{B}-\nabla p
\end{equation}
where the difference between electron and ion velocities just gives us the current density $\textbf{j}$
\begin{equation}\label{current}
    \textbf{j} = ne(\boldsymbol{u}_{i} - \boldsymbol{u}_{e})
\end{equation}
An MPDT which only operates on the internal magnetic field is known as the \textit{Self-field MPDT}. The internal magnetic field comes from the translational motion of electrons leaving cathode. This discharge between the cathode and the cathode results in an azimuthal magnetic field induced by moving electrons. 

If an additional magnetic field is provided externally, powered by the power unit on-board, we would have what we call \textit{Applied-field MPDT} . Such addition of an external high magnetic field provides greater performance and adjustable thrust.

In \ref{fig:enter-label} the direction of magnetic field lines around the radial direction can be seen clearly. Due to current density being radial and the magnetic field azimuthal, resulting Lorentz force is $j_{r} \times \boldsymbol{B}_{\phi}$   

In general, all the thrusters that work upon the principles of plasma follow the momentum equation\cite{alvaro_thesis}:
\begin{equation}\label{plasma_momentumequation}
\nabla \cdot \rho \mathbf{u u}=\epsilon_0(\nabla \cdot \mathbf{E}) \mathbf{E}+\mathbf{j} \times \mathbf{B}-\nabla \cdot \overline{{P}}
\end{equation}

Each term on the right hand-side can be regarded as a different type of an electric thruster. We shall start with the current term $\mathbf{j} \times \mathbf{B}$. Thrusters whose main driving force is the electromagnetic forces are regarded as the \textit{Lorentz thrusters}. In this work, we will not bother with the pressure term $\nabla \cdot \overline{P}$ as this is mostly the working principle of electrothermal engines, which takes advantage of pressure force exerted on nozzle walls.
Similar to Lorentz force term (second equation on the right hand side in equation \ref{plasma_momentumequation}, electrostatic propulsion systems use the electrostatic force, as represented by $\epsilon_0(\nabla \cdot \mathbf{E})\mathbf{E}$. Ion thruster is one such engine.

In Figure \ref{fig:enter-label}, we see a simple schematic of an MPD thruster. The field lines extending from cathode to cathode are the electric field lines created by the voltage difference, where it is fed by a power source, typically in the kW or MW range.

\begin{figure} 
    \centering
    \includegraphics[scale=0.6]{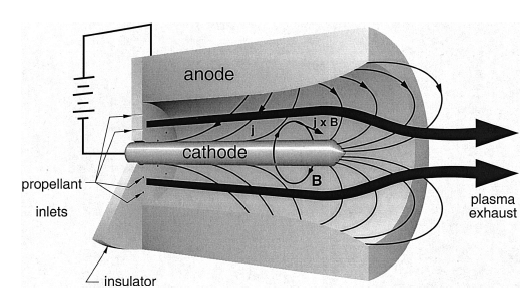}
    \caption{A simple SF-MPD Thruster Schematic}
    \label{fig:somelabel}
\end{figure}

\begin{figure}[!b]
   \begin{minipage}{0.48\textwidth}
     \centering
     \includegraphics[width=.9\linewidth]{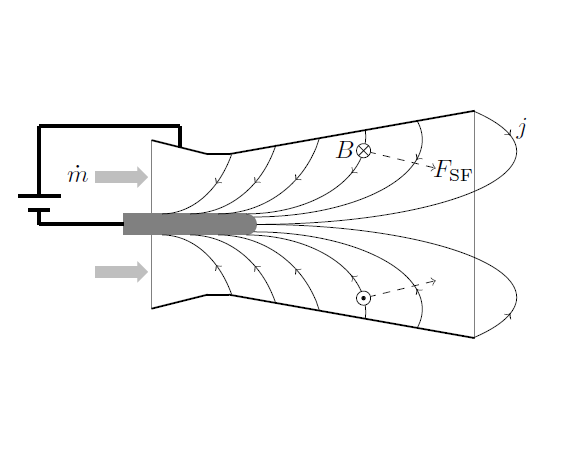}
     \caption{Self-field MPDT}\label{Fig:Data1}\cite{kodys2005critical}
   \end{minipage}\hfill
   \begin{minipage}{0.48\textwidth}
     \centering
     \includegraphics[width=.9\linewidth]{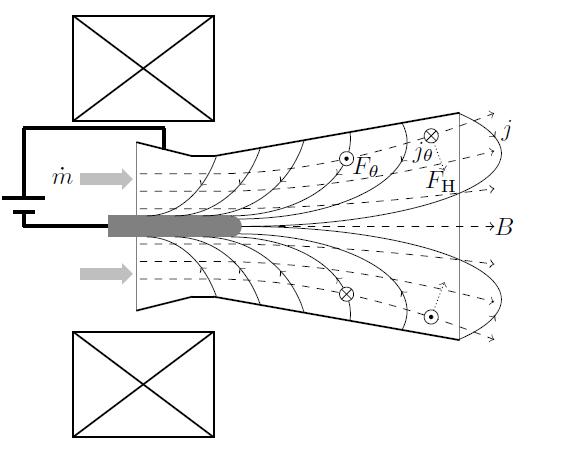}
     \caption{Applied-field MPDT}\label{Fig:Data2}\cite{kodys2005critical}
   \end{minipage}
\end{figure}

\subsection{Step 1- Propellant Injection and Gas Ionization}

An MPD thruster starts by applying the necessary electric field between anode and cathode for ionization. Neglecting the effects of pressure gradients, our primary source of acceleration will be the electric field. The electron energy acquired by this electromagnetic acceleration is the primary source of gas ionization. After leaving cathode, their collisions with inert gasses in the chamber excites the gas molecules, which can rip off electrons from their orbitals, resulting in creation of ions. When the electric field between electrodes are sufficiently high, electrons liberated from cathode may start an electron avalanche.  

\subsubsection{Steady State Current}

In this ideal case, we will study electron ionization up to second degrees(in which, secondary electrons are produced per ionization event). Starting with the primary, the number of electrons reaching anode per second can be given as:
\begin{equation}
    n_d = n_0 \cdot e^{\alpha d}
\end{equation}

Where $\alpha$ is the number of ionising collisions on average, d is the distance from cathode. If this was a steady state process, number of positive ions resulting from collision would equal to number of electrons reaching anode.This would mean that for each electron ejected from cathode would produce 
\begin{equation*}
    \frac{(n_d -n_0)}{n_0}
\end{equation*}

electrons on average, which also equals to number of ions produced. During ionization, secondary mechanism will also take place. Defining 
\begin{itemize}
    \item $\gamma = \text{Number of secondary electrons produced per ionization event at cathode}$
    \item $n_0$ = Number of primary electrons emitted from cathode (per second)
    \item $n_0'$ = Number of secondary electrons emitted from cathode (per second)
    \item $n_0''$ = Total number of electrons leaving cathode (per second)
\end{itemize}
By definition $\gamma$ can be given by:

\begin{equation}
\gamma = \frac{n_0'}{n_0'' \left( e^{\alpha d} - 1 \right)}
\end{equation}
\begin{equation}
n_0'' = \frac{n_0}{1 - \gamma \left( e^{\alpha d} - 1 \right)}
\end{equation}
\begin{equation}
n_d = n_0'' e^{\alpha d} = \frac{n_0 e^{\alpha d}}{1 - \gamma \left( e^{\alpha d} - 1 \right)}
\end{equation}
Steady state current can be given by
\begin{equation}
I = \frac{I_0 e^{\alpha d}}{1 - \gamma \left( e^{\alpha d} - 1 \right)}
\end{equation}
\subsubsection{Saha Ionization Equation}
Although magnetoplasmadynamic thrusters doesn't involve Saha ionization as a direct, primary ionization mechanism, in some variations or designs of MPDT, thermal ionization maybe taken into account as well. If the temperature of the plasma reaches some threshold value where it is sufficient to partially ionize its neighbouring molecules thermally, Saha equation holds. Hence it is worth to discuss here briefly. 

Saha ionization equation relates the temperature and pressure of plasma to degree of ionization when there is a thermal equilibrium and short Debyle length. However There are many degrees of non-equilibirum conditions present, so its application is partially restricted on MPDT. The degree of ionization can be given by\cite{Chen2016}:

\begin{equation}\label{saha_equation}
\frac{n_{i+1} n_e}{n_i} = \frac{2}{\lambda^3} \frac{g_{i+1}}{g_i} \exp \left[ -\frac{(\epsilon_{i+1} - \epsilon_i)}{k_B T} \right]
\end{equation}

Here, $n_i$ represents the density of atoms in the i-th state, $n_e$ is the electron density, $\epsilon$ difference is ionization energy, $g_i$ degeneracy of state, $\lambda$ thermal de Broglie wavelength of electron which is defined as:
\begin{equation}
\lambda^{\text{def}} = \sqrt{\frac{h^2}{2 \pi m_e k_B T}}
\end{equation}

$m_e$ electron mass, \textit{T} temperature, and \textit{h} is Plank's constant.

We can simplify Equation \ref{saha_equation} by just taking one level of ionization into account. This means $n_e$ = $n_1$. And so total density is just $n = n_0 + n_1$. If we label the degree of ionization as $\frac{n_1}{n}=x$, we get

\begin{equation}
\frac{x^2}{1 - x} = A = \frac{2}{n \lambda^3} \frac{g_1}{g_0} \exp \left[ -\frac{\epsilon}{k_B T} \right]
\end{equation}

Resulting in a quadratic equation that if solved for \textit{x}, we acquire the degree of ionization

\begin{equation}
x^2 + A x - A = 0
\end{equation}

\subsubsection{Electric Field, Induced Current and Magnetic Field}

Furthermore, these ions are also subject to the electric field. Due to mass and charge difference, a difference in velocity is observed, which is imparted into current equation, as stated in Equation \ref{current}. Using  the previously defined $\textbf{P}_{ie} = -\textbf{P}_{ei}$, we can rewrite this equation in terms of ion-electron collision frequency $v_{ie}$:
\begin{equation}
    P_{ie} = -P_{ei} = \nu_{ie} m_e n (u_e - u_i) = -\frac{n e}{\sigma} \mathbf{j}
\end{equation}
Here, $\sigma$ represents conductivity, defined as:
\begin{equation}
    \sigma = \frac{n e^2}{\nu_{ie} m_e}
\end{equation}
Substituting them into our previous equations we end up with:
\begin{equation}
    m_i n \frac{d \mathbf{u}_i}{dt} = n e \left( \mathbf{E} + \mathbf{u}_i \times \mathbf{B} \right) - \nabla p_i - \frac{n e}{\sigma} \mathbf{j}
\end{equation}

From here, we may reach the \textit{generalized Ohm's Law}
\begin{equation}
    \mathbf{j} = \sigma \left( \mathbf{E} + \mathbf{u}_e \times \mathbf{B} + \frac{1}{n e} \nabla p_e \right)
\end{equation}
\begin{equation}\label{current_density}
    \mathbf{j} = \sigma \left( \mathbf{E} + \mathbf{u}_i \times \mathbf{B} + \frac{1}{n e} \nabla p_e - \frac{1}{n e} \mathbf{j} \times \mathbf{B} \right)
\end{equation}
From here we can find the electric field vector as:
\begin{equation}
    \mathbf{E} = -\mathbf{u}_i \times \mathbf{B} + \frac{1}{n e} \mathbf{j} \times \mathbf{B} - \frac{1}{n e} \nabla p_e + \frac{\mathbf{j}}{\sigma}
\end{equation}
Each term in this equation has their own characterizations. On the most right-hand side we have our Ohmic term, left to it the pressure gradient, the Hall term, and magnetic force exerted on the ions. It is worth mentioning that some electric propulsion designs use more than a single acceleration mechanism.

In the case of an applied-field MPDT, we can make some assumptions in Equation \ref{current_density} get radial and axial components of current density. Assuming a cylindrical chamber where we have a system similar to current going over a coil, if axial dependencies of electric field and pressure gradient are neglected, we can acquire current density components as\cite{sasoh1992electromagnetic}:
\begin{equation}\label{theta_current}
\mathbf{J}_r = \frac{\overline{\sigma}_0}{1 + (\omega_e \tau_e)^2} \left[ \overline{E}_r + \beta \left( \frac{d p_e}{dr} \right) - (\omega_e \tau_e) u_z \overline{B}_r \right]
\end{equation}

\begin{equation}\label{radial_current}
\mathbf{J}_\theta = \frac{\overline{\sigma}_0}{1 + (\omega_e \tau_e)^2} \left[ (\omega_e \tau_e) \overline{E}_r + \beta (\omega_e \tau_e) \left( \frac{d p_e}{dr} \right) + u_z \overline{B}_r \right]
\end{equation}
where here, $\beta$ represents $\frac{1}{ne}$, $\omega_e \tau_e = \sigma_0 \beta B_z$, and radial component of magnetic field is small compared to axial. Here, we also have the Hall efect term and diamagnetic effect term, represented by first and second terms in bracket respectively. Combining Equation \ref{theta_current} and Equation \ref{radial_current} is pretty straight forward, and will give us a coupled, compact form of current density given as:
\begin{equation}\label{coupled_current}
\mathbf{J}_\theta = (\overline{\omega_e \tau_e}) \mathbf{J}_r + \overline{\sigma}_0 u_z \overline{B}_r
\end{equation}

If we were to add the ionization fraction of plasma into our current Equation \ref{current_density}, we get\cite{sasoh1992electromagnetic}:

\begin{equation}
\mathbf{j} = \sigma \left[ \mathbf{E} + (\mathbf{v} \times \mathbf{B}) \right] - \frac{\Omega}{B} (\mathbf{j} \times \mathbf{B}) + (1 - \alpha) \frac{\Omega \Omega_i}{B^2} \left[ (\mathbf{j} \times \mathbf{B}) \times \mathbf{B} \right]
\end{equation}

This time, ionization fraction $\alpha$ and electron Hall parameter $\Omega$, which is product of electron cyclotron frequency and electron collision time, ion Hall parameter $\Omega_i$ (analogous to $\Omega$, but defined in terms of ion-neutral collision time and ion cyclotron frequency also takes part.

Ohm's law can be modified to find:

\begin{equation}\label{modified_ohmslaw}
\mathbf{E} = -\nabla \phi = \frac{1}{\sigma} \left[ \mathbf{j} + \frac{\Omega}{B} (\mathbf{j} \times \mathbf{B}) \right] - (\mathbf{v} \times \mathbf{B})
\end{equation}

If one wants to find potential drop across plasma, equation \ref{modified_ohmslaw} can be integrated over the space accordingly.

The magnetic transport equation can be found in the same manner. Starting with generalized Ohm's law and combining it with Maxwell equation:
\begin{equation}
\nabla \times \mathbf{E} = -\frac{\partial \mathbf{B}}{\partial t}
\end{equation}
We can get:
\begin{equation}
= \nabla \times \left( \frac{\nabla \times \mathbf{B}}{\mu_0 \sigma} \right) + \nabla \times \left[ \frac{\Omega}{\mu_0 \sigma B} ((\nabla \times \mathbf{B}) \times \mathbf{B}) \right] - \nabla \times (\mathbf{v} \times \mathbf{B})
\end{equation}

This equation above will give the result of induced magnetic field distribution. When dealing with an applied field MPDT, Maxwell's equations solutions for external current distribution will be taken into account.

\subsection{Step 2 - Force Vector}
\subsubsection{Force in Self-Field MPDT}

Usually, MPDT takes advantage of both Lorentz force, and electrothermal (pressure) force. As addition of electrothermal term may complicate our calculations, for this section we will assume we have a high efficiency MPDT which works solely based on electromagnetic means. For an azimuthally induced magnetic field $\mathbf{B} =B_0\hat{\theta}$ in a SF-MPDT, force in the chamber can be given as\cite{alvaro_thesis}:
\begin{equation}
\mathbf{F}_{EM} = \iiint_V \mathbf{j} \times \mathbf{B} \, dV = \frac{1}{\mu_0} \iiint_V (\nabla \times \mathbf{B}) \times \mathbf{B} \, dV
\end{equation}
Where we substitute the vector identity
\begin{equation}
(\nabla \times \mathbf{B}) \times \mathbf{B} = \mathbf{B} (\nabla \cdot \mathbf{B}) - \nabla \left( \frac{B^2}{2} \right)
\end{equation}
Force is then just:
\begin{equation}
\mathbf{F}_{EM} = \frac{1}{\mu_0} \iiint_V \mathbf{B} (\nabla \cdot \mathbf{B}) \, dV - \frac{1}{\mu_0} \iiint_V \nabla \left( \frac{B^2}{2} \right) dV
\end{equation}
\begin{equation}
= -\frac{1}{\mu_0} \iiint_V \left( \frac{B_\theta^2}{r} \mathbf{i}_r + \nabla \left( \frac{B_\theta^2}{2} \right) \right) dV
\end{equation}
This result implies that the force vector has two components, namely axial and radial. Breaking integral into two parts we see:

\begin{equation}
F_r = -\frac{1}{\mu_0} \iiint_V \left( \frac{B_\theta^2}{r} + \frac{\partial}{\partial r} \frac{B_\theta^2}{2} \right) dV = -\frac{1}{\mu_0} \iiint_V \frac{1}{r^2} \frac{\partial}{\partial r} \frac{r^2 B_\theta^2}{2} dV \tag{3.24}
\end{equation}
for radial and
\begin{equation}
F_z = -\frac{1}{\mu_0} \iint_V \frac{\partial}{\partial z} \frac{B_\theta^2}{2} dz dA \tag{3.25}
\end{equation}
for axial component. Assuming the electrons do a uniform cylindrical motion, magnetic field can be approximated to be the same as a current flowing in a coil\cite{Maecker1955}:
\begin{equation}
B_\theta = \frac{\mu_0 I_d}{2 \pi r}
\end{equation}
\begin{equation}
F_z = \frac{\pi}{\mu_0} \int_{r_c}^{r_a} \left( \frac{\mu_0 I_d}{2 \pi r} \right)^2 r \, dr = \frac{\mu_0 I_d^2}{4 \pi} \ln \left( \frac{r_a}{r_c} \right) \tag{3.27}
\end{equation}
\subsubsection{Force in Applied-Field MPDT}
Similarly, when dealing with an applied-field MPDT, we can make use of Equation \ref{theta_current} and Equation \ref{radial_current} to obtain electromagnetic force. Defining acceleration region as the inside of a coil, thrust will be just the integration of Lorentz force inside the nozzle, given as\cite{sasoh1992electromagnetic}:
\begin{equation}
F_{em} = \dot{m} u_z
\end{equation}
\begin{equation}
= \int_{0}^{R} \left[ (-j_\theta) B_r L + \int_{r}^{R} (-j_\theta) B_z dr' \right] 2\pi r dr
\end{equation}
\begin{equation}
= V_{eq} (-\overline{j}_\theta) B_0
\end{equation}
Here $V_eq$ represents equivalent volume of nozzle and $B_0$ is the applied magnetic field strength. Note that we now have also used mass flow rate in force equation. Using current density we have found earlier (\ref{coupled_current}) 
\begin{equation}
\overline{J}_\theta = \frac{ \left[ \left( \frac{B^2}{\dot{m}} \right)_c \frac{B^2}{\dot{m}} \right]^{1/2} }{ \frac{B^2}{\dot{m}} + \left( \frac{B^2}{\dot{m}} \right)_c } \overline{J}_{\theta,c}
\end{equation}
Where
\begin{equation}
\left( \frac{B^2}{\dot{m}} \right)_c = \frac{1}{\phi V_{eq} \overline{\sigma}_0}
\end{equation}
\begin{equation}
\overline{J}_{\theta,c} = \left( \frac{\overline{\sigma}_0 \dot{m}}{\phi V_{eq}} \right)^{1/2} \beta \overline{J}_r = (\omega_e \tau_e)_c \overline{J}_r
\end{equation}
Azimuthal current density gets its highest value when $\frac{B^2}{\dot{m}} = \left( \frac{B^2}{\dot{m}} \right)_c$. Combining all we have found:
\begin{equation}
F_{em} = \frac{\frac{B^2}{\dot{m}}}{\frac{B^2}{\dot{m}} + \left( \frac{B^2}{\dot{m}} \right)_c} F_{em,max}
\end{equation}
where
\begin{equation}
F_{em,max} = \frac{\beta (-\overline{J}_r) \dot{m}}{\phi} = (\omega_e \tau_e)_c (-\overline{J}_r) B_c V_{eq}
\end{equation}
\begin{equation}
B_c = \left( \frac{\dot{m}}{\phi V_{eq} \overline{\sigma}_0} \right)^{1/2}
\end{equation}
After finding force, we can relate it to specific impulse $I_{sp}$ as
\begin{equation}
I_{sp} = \frac{\frac{B^2}{\dot{m}}}{\frac{B^2}{\dot{m}} + \left( \frac{B^2}{\dot{m}} \right)_c} I_{sp,max}
\end{equation}
where
\begin{equation}
I_{sp,max} = \frac{\beta (-\overline{J}_r)}{\phi g}
\end{equation}
\subsection{Plasma Detachment}

Now that the formation and guidance of plasma is complete, it is now time to divert the plasma exhaust in such a way that it will obtain an efficient thrust. Decoupling of plasma from magnetic field lines is done by using divergence of magnetic field lines and the inertia of plasma. The key aspect of this detachment process is regarded as Alfvenic flow. Transition from sub-Alfvenic to super-Alfvenic ensures the average kinetic energy of plasma exceeds energy density of magnetic field. This means that divergence of magnetic field causes field lines to stretch and become wide open at the end of the nozzle. Through this transition, plasma can move independently of the magnetic field. Smooth transition has to be ensured otherwise the plasma can get restrained in magnetic field, hence compromising thrust.

Plasma release condition is based on magnetic pressure and plasma dynamic pressure where the relation can be given by:
\begin{equation}
    \frac{B^2}{2\mu_0} \le \rho v^2
\end{equation}
In terms of Alfvén speed $v_A$,

\begin{equation}
    v_{A} = \frac{B}{\sqrt{\mu_0 \rho}}
\end{equation}

A plasma flow is considered sub-Alfvenic if magnitude of flow velocity is less than this value. This implies that plasma is purely guided and dominated by magnetic field. Any velocity that exceeds this value causes plasma to detach, and no longer follow field lines.

When the nozzle has a conical geometry, the magnetic field exhibits monopolelike configuration. In absence of plasma (i.e) magnetic field components are shown as\cite{plasma_Detachment}:

\begin{equation}
B_r = \frac{r \Phi_0}{(1 - \cos \theta_0) (r^2 + z^2)^{3/2}}
\end{equation}

\begin{equation}
B_z = \frac{z \Phi_0}{(1 - \cos \theta_0) (r^2 + z^2)^{3/2}}
\end{equation}
Where $\Phi$ represents the magnetic flux function.
The radial and axial component equation of magnetic field is also satisfied in presence of plasma. This means this are the equivalent magnetic fields in a state state condition. Relating magnetic flux and radial position as:
\begin{equation}
r(\Phi_0, z) = z \tan \theta_0,
\end{equation}
\begin{equation}
r(\Phi, z) = z \sqrt{\left(1 - \left(1 - \cos \theta_0\right)\frac{\Phi}{\Phi_0}\right)^{-2} - 1}.
\end{equation}
\begin{equation*}
\eta \equiv \frac{\dot{P}_z^{\text{out}}}{\dot{P}_z^{\text{in}}}.
\end{equation*}
If we assume that flow velocity and plasma density stays constant throughout the incoming flow, the nozzle efficiency of engine can be related to nozzle divergence angle as\cite{plasma_Detachment}:
\begin{equation*}
\eta \approx 1 - \frac{\theta_0^2}{4}.
\end{equation*}
\begin{figure}[t]
    \centering
    \includegraphics[width=0.6\textwidth]{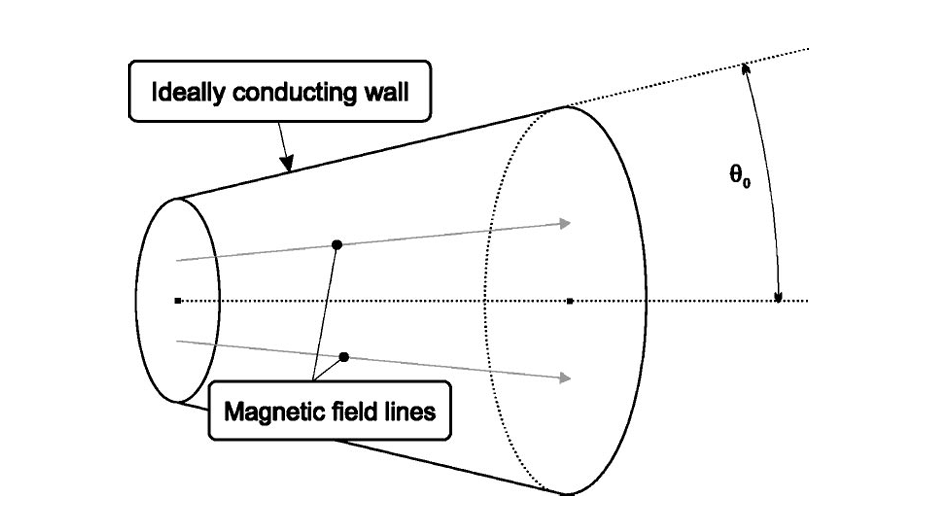}
    \caption{A sample conical nozzle. Nozzle divergence angle is shown as $\theta$\cite{plasma_Detachment}}
    \label{fig:enter-label}
\end{figure}
\begin{figure}[h]
    \centering
    \includegraphics[width=0.5\textwidth]{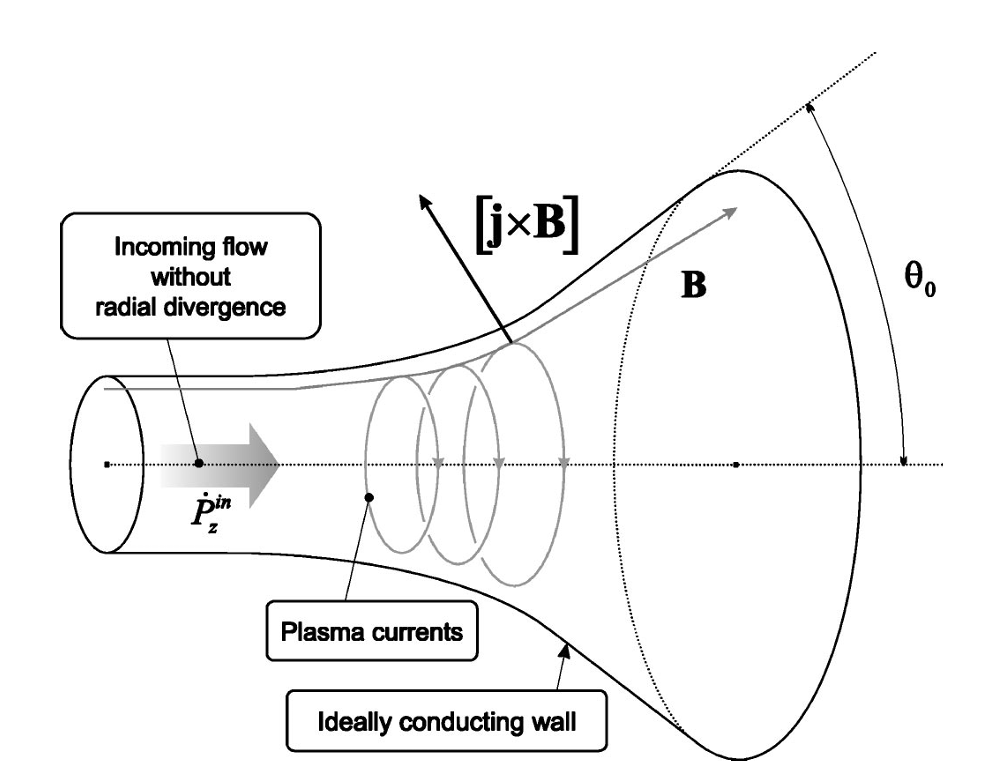}
    \caption{Nozzle curvature can be seen here more clearly.\cite{plasma_Detachment}}
    \label{fig:enter-label}
\end{figure}

\section{Conclusions}
In this report, we have discussed types of electric propulsion systems, and saw how they differ from chemical engines. We have seen the main (and mostly ideal) equations governing a standard magnetoplasmadynamic thruster and application of magneto hydrodynamic flow equations, as well as its two variations, namely applied-field and self-field. The characteristics of such engine discussed here promises an efficient and cost-effective thruster technology that is suitable for deep space missions, satelite orbit corrections, crewed missions and even cargo ships on oceans. In summary,

\begin{enumerate}
    \item \textbf{\textit{Specific Impulse}}: The ability of MPDT to provide a decent level of specific impulse is an essential asset for different mission profiles.
    \item \textbf{\textit{Ion-Cylotron Collision Ionization}} : Using electrons as primary source of ionization, the engine ionizes and heats the propellant gas. This type of ionization is incredibly efficient in terms of plasma creation.
    \item \textbf{\textit{Self and Applied Field Equations}} : Direction and magnitude of current density, thrust, and magnetic field were studied in each variation on engine.
    \item \textbf{\textit{Advantages Over Chemical Propulsion}}: MPDT offers several advantages over traditional chemical rockets, including better efficiency, lower mass of propellant, and the ability to provide continuous thrust. These features could shine on deep space missions that require extended periods of thrust.
    \item \textbf{\textit{Challenges and Future Work}} : Departing from its advantages, MPDT is troubled with the necessity for ample electrical power and advanced thermal management systems. Ongoing research and development are going on on this issue to optimize the overall system reliability and performance.
\end{enumerate}

Although everything listed in this work discusses the pros of the engine, MPDT is still not mature enough and currently under heavy development. The drawbacks of this technology are the heavy power dependency of the engine (which can be in the order of MWs), complexity of magnetic fields and thermal management. It is still not yet clear enough on how the MPDT will supply its high power demanding components. Nuclear reactors and solar arrays are currently on the table but use of reactors on a crewed mission is regulated and currently prohibited.

With this regard, MPDT is a promosing technology for faster and more efficient travels. This is necessary for further development and space demonstration exercises to realize the full potential of MPDT for crewed and uncrewed missions. This conclusion summarizes the detailed analysis provided in the paper on the current status and future potential of MPDT technology.

\begin{figure}[!htb]
   \begin{minipage}{0.5\textwidth}
     \centering
     \includegraphics[width=.7\linewidth]{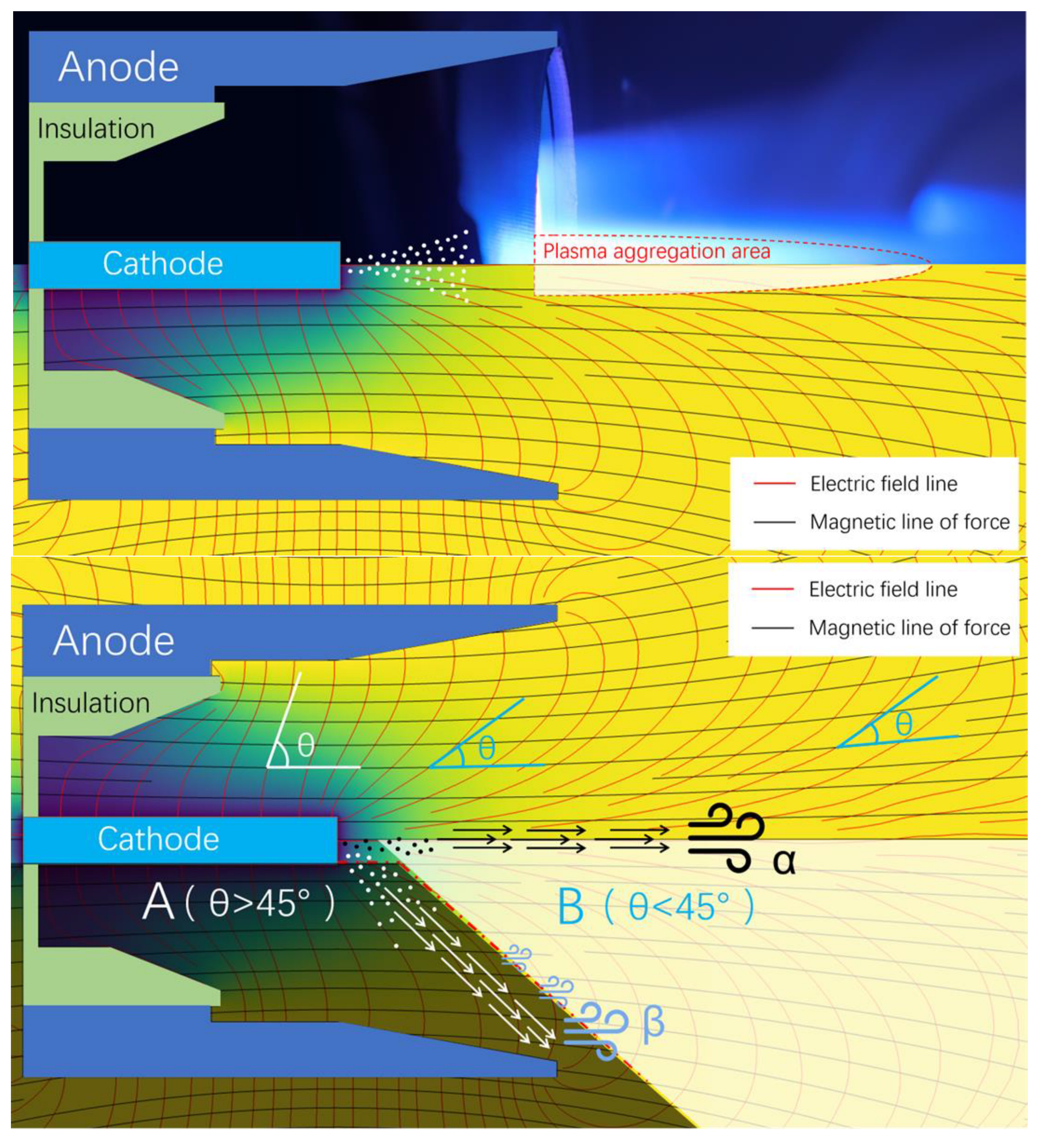}
     \caption{3D Electric and Magnetic Field Model\cite{aerospace10020124}}\label{Fig:Data3}
   \end{minipage}\hfill
   \begin{minipage}{0.6\textwidth}
     \centering
     \includegraphics[width=.7\linewidth]{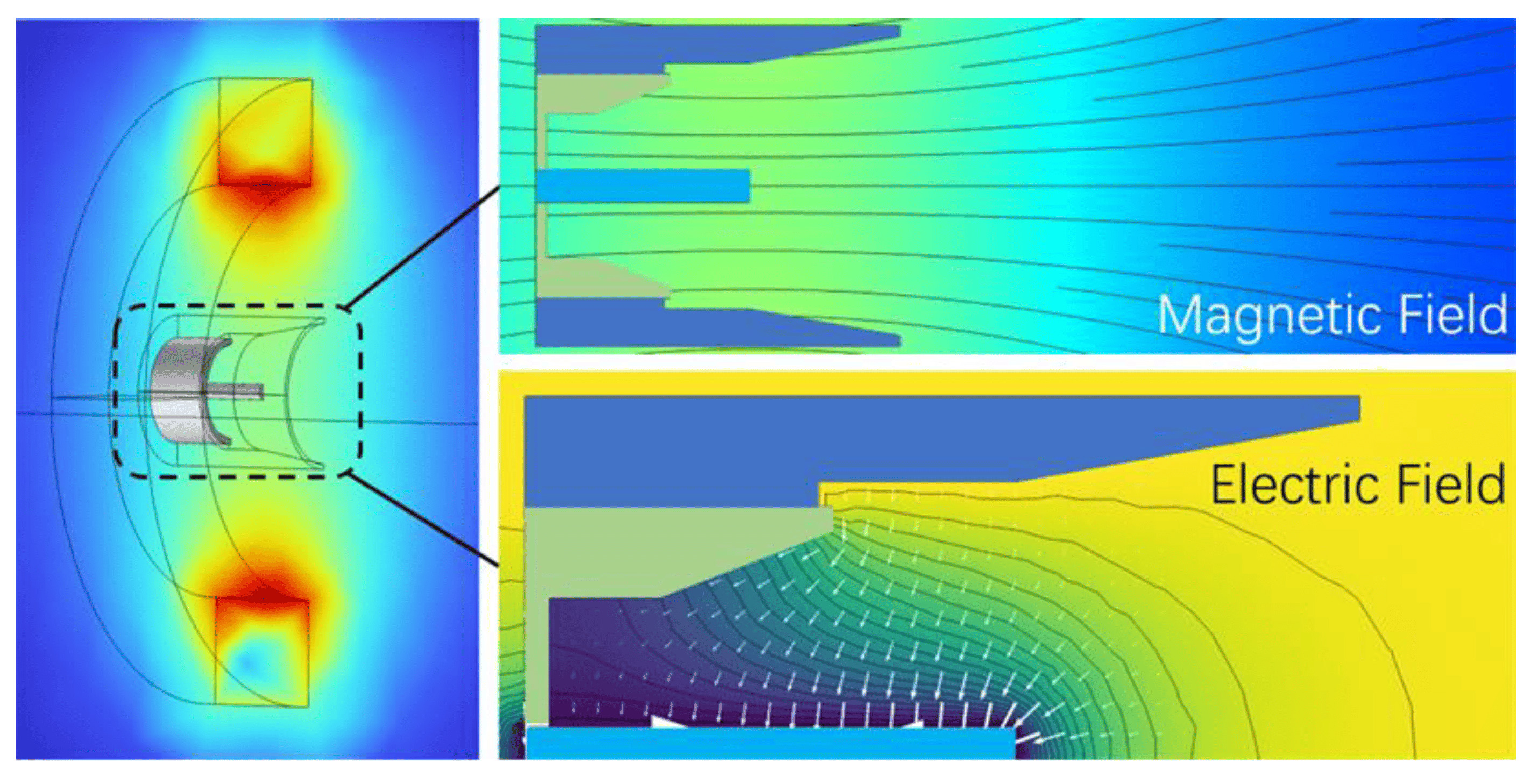}
     \caption{3D Electric and Magnetic Field Model\cite{aerospace10020124}}\label{Fig:Data4}
   \end{minipage}
\end{figure}

\newpage

\bibliographystyle{ieeetr}
\bibliography{sample}

\end{document}